\documentclass[journal,comsoc]{IEEEtran}

\usepackage[T1]{fontenc}

\usepackage{hyperref}
\usepackage{hypcap}

\usepackage{xcolor}
\ifCLASSINFOpdf

\else

\fi

\usepackage{float}
\usepackage{multirow}
\usepackage{colortbl}
\usepackage{makecell}
\usepackage{amsmath}

\usepackage[cmintegrals]{newtxmath}

 \usepackage{graphicx}

\begin{document}

\title{Towards Self-Powered Internet of Underwater Things Devices}
\author{Authors
\author{Jose Ilton de Oliveira Filho,~\IEEEmembership{Student Member,~IEEE,}
Abderrahmen Trichili,~\IEEEmembership{Member,~IEEE,}
Boon S. Ooi,~\IEEEmembership{Senior Member,~IEEE}
Mohamed\nobreakdash-Slim Alouini,~\IEEEmembership{Fellow,~IEEE,}
and Khaled Nabil Salama,~\IEEEmembership{Senior Member,~IEEE}

\thanks{Authors are with the Computer, Electrical and Mathematical Sciences $\&$ Engineering in King Abdullah University of Science and Technology, Thuwal, Makkah Province, Kingdom of Saudi Arabia.}
}

\thanks{Manuscript received June, 2019}}

\markboth{IEEE Communications Magazine}%
{Shell \MakeLowercase{\textit{et al.}}: Bare Demo of IEEEtran.cls for IEEE Communications Society Journals}

\maketitle

\begin{abstract}
Exploiting light beams to carry information and deliver power is mooted as a potential technology to recharge batteries of future generation Internet of things (IoT) and Internet of underwater things (IoUT) devices while providing optical connectivity. Simultaneous lightwave information and power transfer (SLIPT) has been recently proposed as an efficient way for wireless power transfer between communicating terminals. In this article, we provide an overview of the various SLIPT techniques in time, power, and space domains. We additionally demonstrate two SLPIT scenarios through underwater channels. Moreover, we discuss the open issues related to the hardware as well as system deployment in harsh environments.
\end{abstract}
\IEEEpeerreviewmaketitle

\section{Introduction}
\IEEEPARstart{W}{ireless} power transfer has received a considerable attention in recent years. Radio frequency (RF) simultaneous wireless information and power transfer (SWIPT) was proposed as a technique to transmit information and harvest energy by converting energy from an electromagnetic field into the electrical domain \cite{RFSWIPT1,RFSWIPT2}. SWIPT is considered as a future generation energy transfer technology in wireless communication networks. However, besides the RF spectrum scarcity, RF energy harvesting suffers from relatively low efficiency and major technical problems related to the transmitting, and receiving circuits \cite{RFSWIPT3}. Additional challenges are imposed by the electromagnetic safety and health concerns raised over the high power RF applications (references are within \cite{RFSWIPT3}). SWIPT can be equally a source of interference to data transmission, and RF pollution.\newline  
\indent One alternative to the use of electromagnetic radiation to harvest energy is lightwaves emitted by light emitting diodes (LEDs) and lasers sources. Using light beams, it is possible to simultaneously perform a transfer of energy and efficiently deliver data streams. Simultaneous lightwave information and power transfer (SLIPT) can provide significant performance compared to RF-based SWIPT taking the advantage of the free-license optical wireless technology \cite{Georges1}.  Lightwave energy harvesting can be also a complementary technology to visible light communication (VLC) as proposed in \cite{VLCRFSWIPT,SLIPT,SLIPTVLC}. In a dual-hop VLC/RF configuration, Rakia \textit{et al.} proposed harvesting energy from the VLC hop, by extracting the direct current (DC) component of the receiver illuminated by an LED \cite{VLCRFSWIPT}. The harvested energy is then harnessed to re-transmit the information over the RF link. Authors of \cite{SLIPT}, demonstrated the use of a low-cost Silicon solar cell to decode  low-frequency VLC signals and harvest optical energy over a short propagation distance of 40 cm. In \cite{SLIPTVLC}, authors proposed the use of VLC systems with energy harvesting capabilities for night gathering events. Diamantoulakis \textit{et al.} proposed new strategies to enhance the efficiency and optimize between the communication and the energy harvesting functions for indoor applications through visible and infrared wireless communications.\newline
\begin{figure}[htp]
\centering
\includegraphics[width=3.5in]{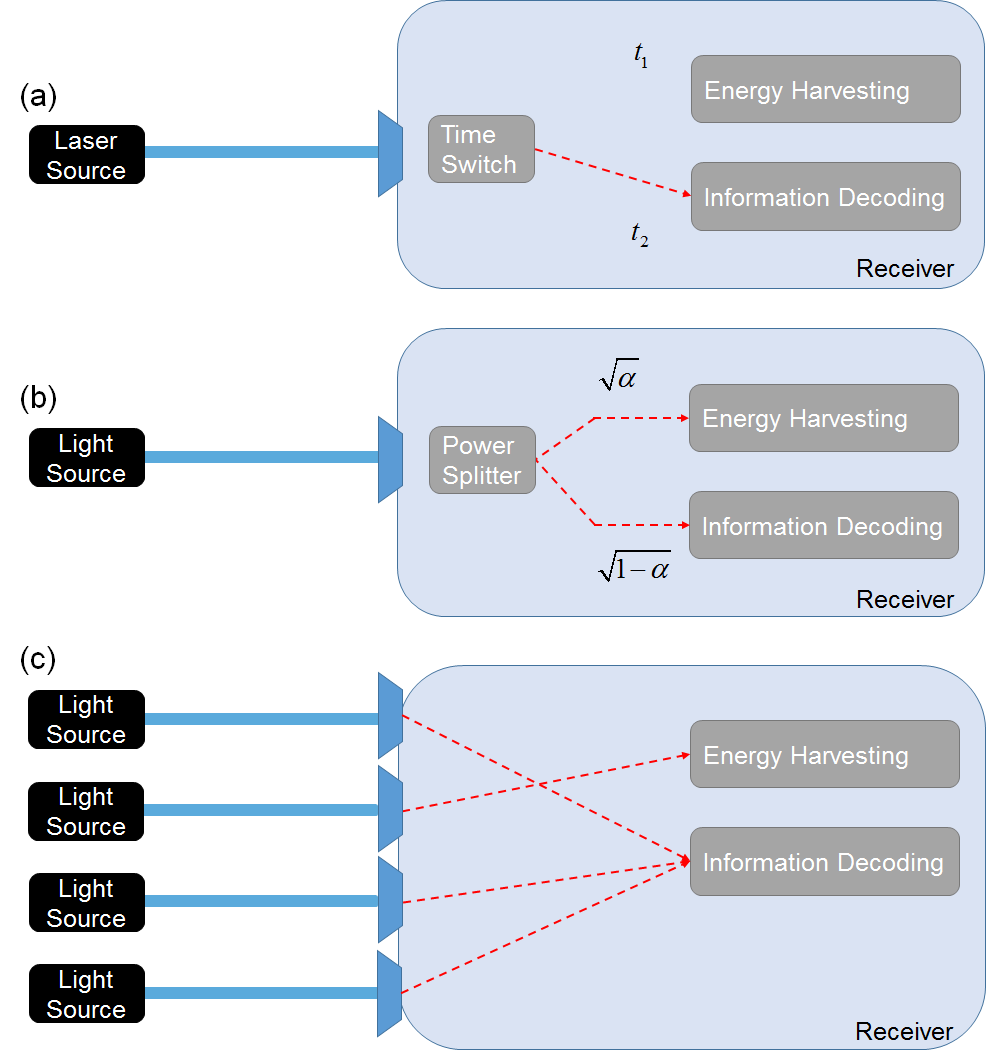}
\caption{Different SLIPT techniques: (a) time switching, (b) power splitting, and (c) spatial splitting.}
\label{SLIPTTech}
\end{figure}

\begin{figure*}[!t]
\centering
\includegraphics[width=6in]{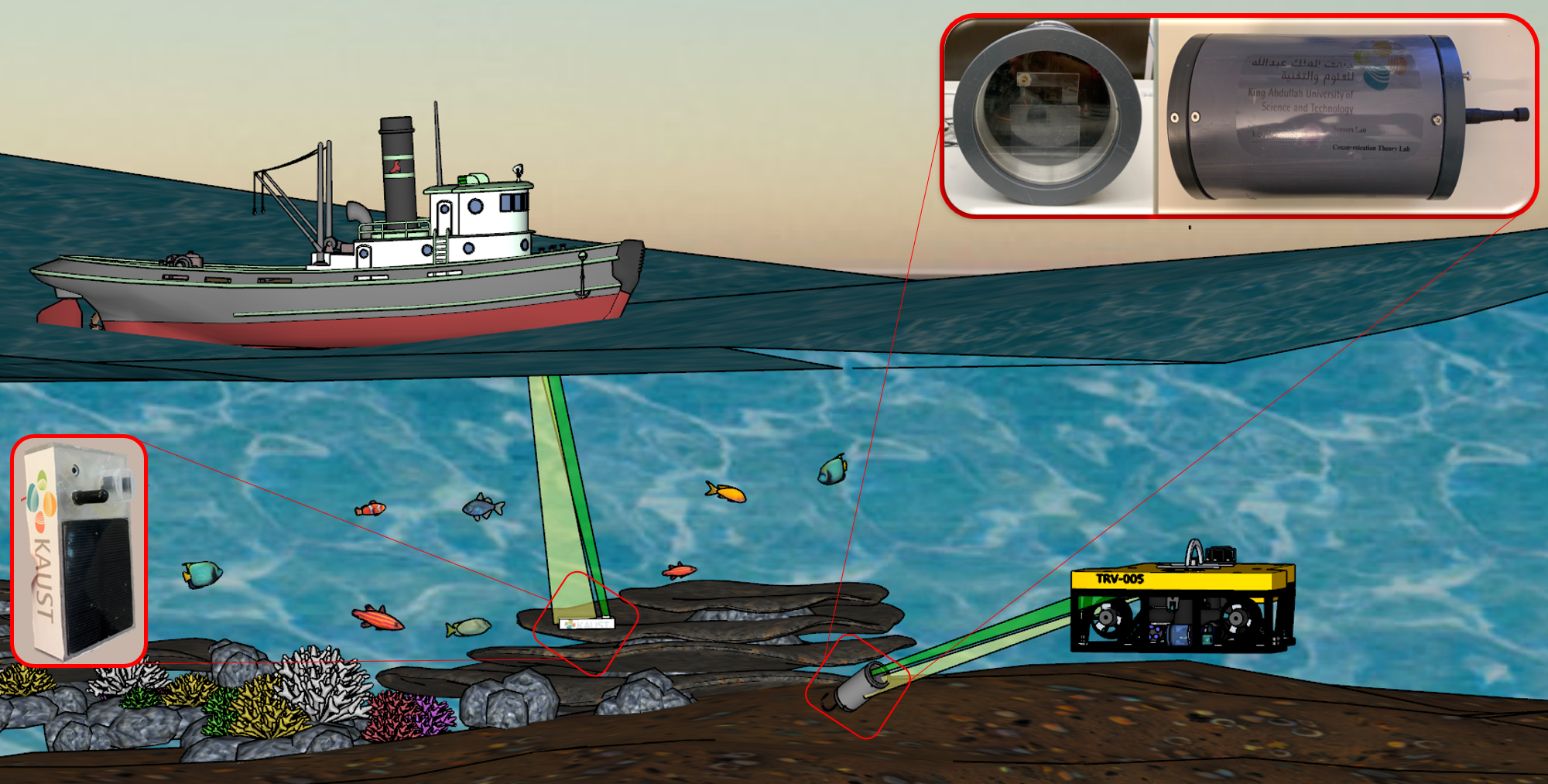}
\caption{Illustration of the use of self-powered IoUT devices in an underwater environment.}
\label{UnderwaterSLIPT}
\end{figure*}

\indent Motivated by the low-cost of the optical communication/energy harvesting circuits components compared to their RF counterparts, SLIPT can be equally a cost effective solution for energy-constrained wireless systems including remote sensors, autonomous self-powered devices. SLIPT can be also very promising for applications in RF-sensitive environment and applications such as medical, smart houses, and aerospace \cite{Georges2}.\newline
Another potential application of SLIPT is powering Internet of underwater things (IoUT) devices. This is motivated by the high demands for underwater communication systems due to the on-going expansion of human activities in underwater environments such as marine life monitoring, pollution control/tracking, marine current power grid connections, underwater exploration, scientific data collection, undersea earthquakes monitoring, and offshore oil field exploration/monitoring. \newline
Wireless transmission under water can be achieved through radio, acoustic, or optical waves. 
Traditionally, acoustic communication has been used for underwater applications for more than 50 years and can cover long ranges up to several Kilometers. The typical frequencies associated with this communication type are 10 Hz to 1 MHz. However, it is well known that this technology suffers from a very small available bandwidth, and large latencies due to the low propagation speed. RF waves suffer from high attenuation in sea-water and can propagate over long distances only at extra low frequencies (30-300 Hz). This requires large antennas and high transmission powers which make them unappealing for most practical purposes. Compared to acoustic and RF communications, underwater optical wireless communication (UWOC) through ocean water can covers large bandwidth and involve transmitting high data amount in the order of sevral Gbit/s. SLIPT as a complementary technology to UWOC can provide continuous connectivity and wireless powering for devices and sensors in difficult-access locations. \newline
\indent In this paper, we present the different concepts of optical SWIPT or SLIPT. We then provide experimental demonstrations of time switching SPLIT in an underwater environment for IoUT applications.  We further discuss the open problems and propose key solutions for the deployment of underwater SLIPT-based devices.


\section{SLIPT System Design and Concept}
Different possible techniques in various domains including, time, power, and space can be adopted to transmit information and harvest energy.
\subsection{Time Switching}
In a time switching configuration, the receiver, which is possibly a low-cost solar cell, switches between the energy harvesting mode and the information decoding mode, better known as the photovoltaic and the photoconductive modes, respectively. Both SLIPT functions are performed over two different time slots $t_{1}$ and $t_{2}$, as can be seen in Fig.~\ref{SLIPTTech} (a). The quantity of harvested energy is ruled by the conversion efficiency of the solar cell. With the advances in the development of high-speed light sources, the maximum transmission rate that could be achieved is mainly restricted by the bandwidth of the solar cell. Synchronizing the photovoltaic mode and the photoconductive mode is crucial and can be done via hardware programming. The switching function can be fulfilled via a low-power relay. 

\begin{figure*}[htp!]
\centering
\includegraphics[width=6.5in]{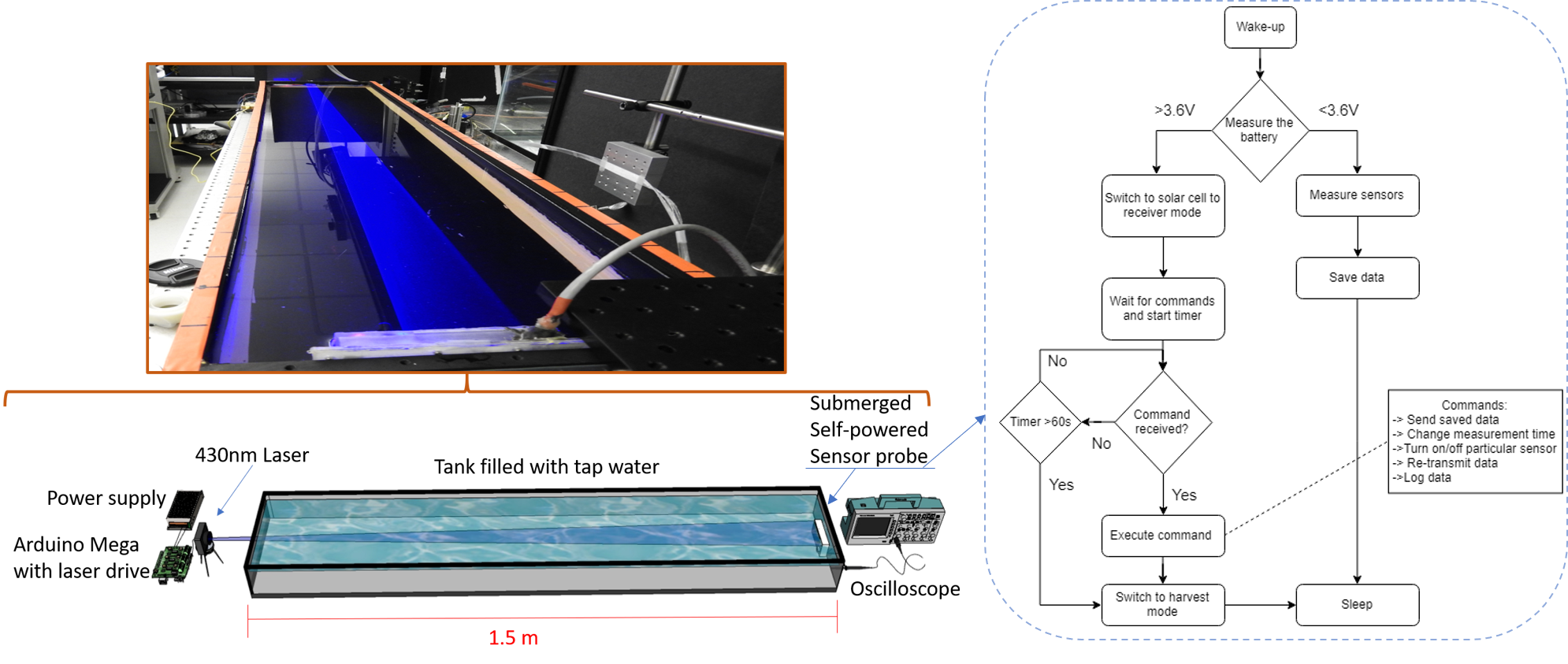}
\caption{ Schematic illustrating the experimental setup of time switching underwater SLIPT and its block diagram program.}
\label{UnderwaterSetup}
\end{figure*}

\subsection{Power Splitting}
Within the power splitting approach, the receiving terminal is simultaneously charged while decoding information carried by the incident light beam, as can be seen in Fig.~\ref{SLIPTTech} (b). A key device needed for this configuration is the power splitter, which splits the incident power into $\alpha$ and $(1-\alpha)$ quantities. The $\alpha P_{R}$ power portion is used to harvest energy, while the $(1-\alpha)P_{R}$ portion is used to decode the received signal. The power splitting component can be either a passive beam splitter, which splits the incident beam from a light source into two or more beams, with  (un)evenly fixed distributed power portions. It can be also a splitter with variable splitting ratios, which can potentially increase the system complexity.  Differently from the time switching approach, using power splitting, the simultaneous energy and power transfer is fulfilled. With this method, it is also possible to achieve higher transmission rates because the decoding can be performed via a high-speed photodiode (PD).

\subsection{Spatial Splitting }
The spatial splitting approach is applied on a configuration involving multiple transmitters and multiple receivers with information decoding and energy harvesting capabilities, as depicted in Fig.~\ref{SLIPTTech} (c). Each transmitter can  transfer data or energy, and each receiver can harvest energy from multiple transmitters. Time switching can be applied within this configuration where the same receiver can act as an ``energy harvester'' and ``signal decoder'' over different time slots.

\section{Demonstrations}

Multiple theoretical and experimental SLIPT-related studies, in free space and underwater media, have been proposed in the literature. Wang \textit{et al.} proposed a novel design of a solar panel as a photo-receiver \cite{Demonstration1}. Authors of \cite{Demonstration2} established a VLC link and used an organic solar panel as a receiver. Earlier demonstrations also involved the use of a 5 cm$^2$ solar panel as a receiver for an underwater communication link over a 7 m long water tank \cite{Demonstration3}. A Gallium Arsenide (GaAs) solar cell was further used to perform a 0.5 Gb/s transmission over a 2 m-long free space link \cite{Demonstration4}. Here, we demonstrate two communication and energy harvesting scenarios over underwater media using two devices, which can be potentially charged through light beams emitted from a source fixed on a boat or an autonomous underwater vehicle (AUV) that could be used in real-life marine environmental monitoring and scientific data collection applications, as depicted in Fig. \ref{UnderwaterSLIPT}. In a first experiment, we charge the battery of a submerged module with temperature and turbidity sensors, and transmit commands using a single laser. The temperature sensor is then used to monitor the variable temperature of a water tank. In a second demonstration, we report charging the capacitor of an IoUT device equipped with a camera and a low-power laser for real-time video streaming.


\subsection{Self-Powered Sensor Module}
\begin{figure}[htp!]
\centering
\includegraphics[width=3.5in]{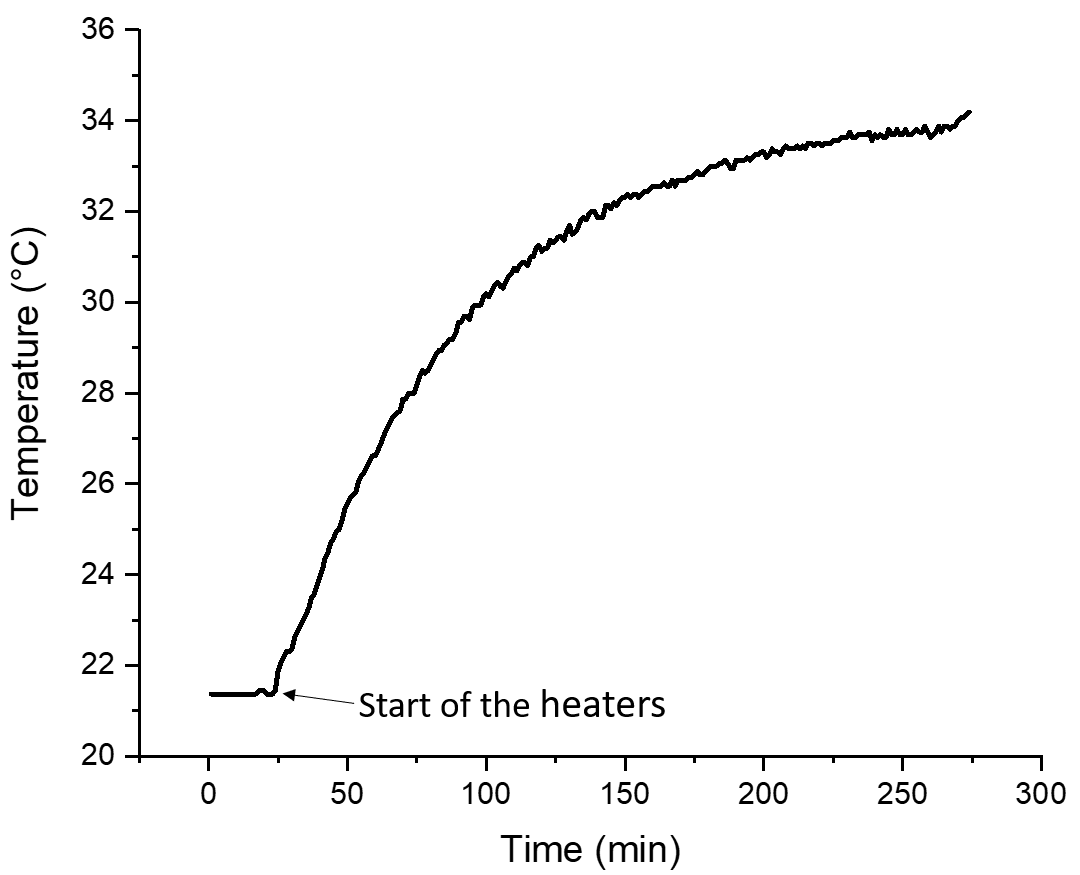}
\caption{Temperature evolution of the water tank as a function of time.}
\label{UnderwaterTemp}
\end{figure}

The experimental setup of the first demonstration is depicted in Fig.~\ref{UnderwaterSetup}. The transmitter is composed by an Arduino Mega with a laser drive connected to a 430 nm laser diode (LD), and is fixed outside of a 1.5-m-long water tank. The receiver, a self-powered sensor platform formed by a a $55\times70$ mm solar cell, with a 3-dB bandwidth of 30 KHz in photovoltaic mode, and an electric circuitry, is placed inside the water tank. The state of the solar cell is controlled via a low-power relay, which changes the connection of the solar cell to the circuit. The solar cell can then have two possible states:
\begin{itemize}
\item It directly deliver power to a battery or a super-capacitor, at the same time its charge is monitored by a wake-up circuit.
\item It is reverse-voltage-biased and the output current passes through a transimpedance amplifier and a comparator. Both circuits are implemented by a programmable system on-chip (PSoC). The main circuit is connected to a low-power microcontroller that receives the signal and processes the data (saving the data or executing the commands). 
\end{itemize}
We should stress that in some of the previously demonstrated SLIPT circuits, such as the one in \cite{Demonstration3}, the harvested energy and the data rate are directly associated to the frequency of the electrical signal due to the coupling capacitor and inductor. Moreover, the feasibility of incorporating a maximum power point tracking (MPPT) for the solar cells with the existing circuit methodology has not been demonstrated before to the best of our knowledge. Through our circuit design, in the energy harvesting mode, we allow the implementation of any MPPT circuit without affecting the data-transmission, as both solar cell modes are independent.  We also note that one of the major challenges for the deployment of UWOC systems  is fulfilling the pointing, acquisition and tracking (PAT) requirements. Using the solar cell as a photo-detector can reduce the strict PAT requirements.
As depicted in Fig.~\ref{UnderwaterSetup}, the module wake-up upon receiving a light beam on the solar panel. The battery voltage $V_{B}$ is then  measured. If $V_{B}<V_{th}$, where $V_{th}$ is a threshold voltage of 3.6V for our module, the sensors start measuring turbidity and/or temperature. The collected data is saved on a secure digital (SD) card fixed on the module, which then enters to a sleep mode. If $V_{B}\ge V_{th}$, the solar panel switches to a receiver mode and receive commands, which include switching on/off a particular sensor, and sending/re-transmitting the saved data on the memory card. Upon executing all the needed commands, the solar panel switches to energy harvesting mode and when reaching a full battery charge, the module enters to the sleep mode. The full charge of the 840 mW module battery takes approximately 124 mins using the blue laser and the reached throughput when the solar panel acts as an information receiver is equal to 500 kbit/s. The temperature sensor is then switched on  to measure the temperature inside the water tank over a time window of more than two hours. The water temperature is controlled using two chillers fixed on the two sides of the tank. The temperature evolution as a function of time is shown in Fig.~\ref{UnderwaterTemp}.
\subsection{Self-Powered Underwater Camera}
\begin{figure}[htp!]
\centering
\includegraphics[width=3.5in]{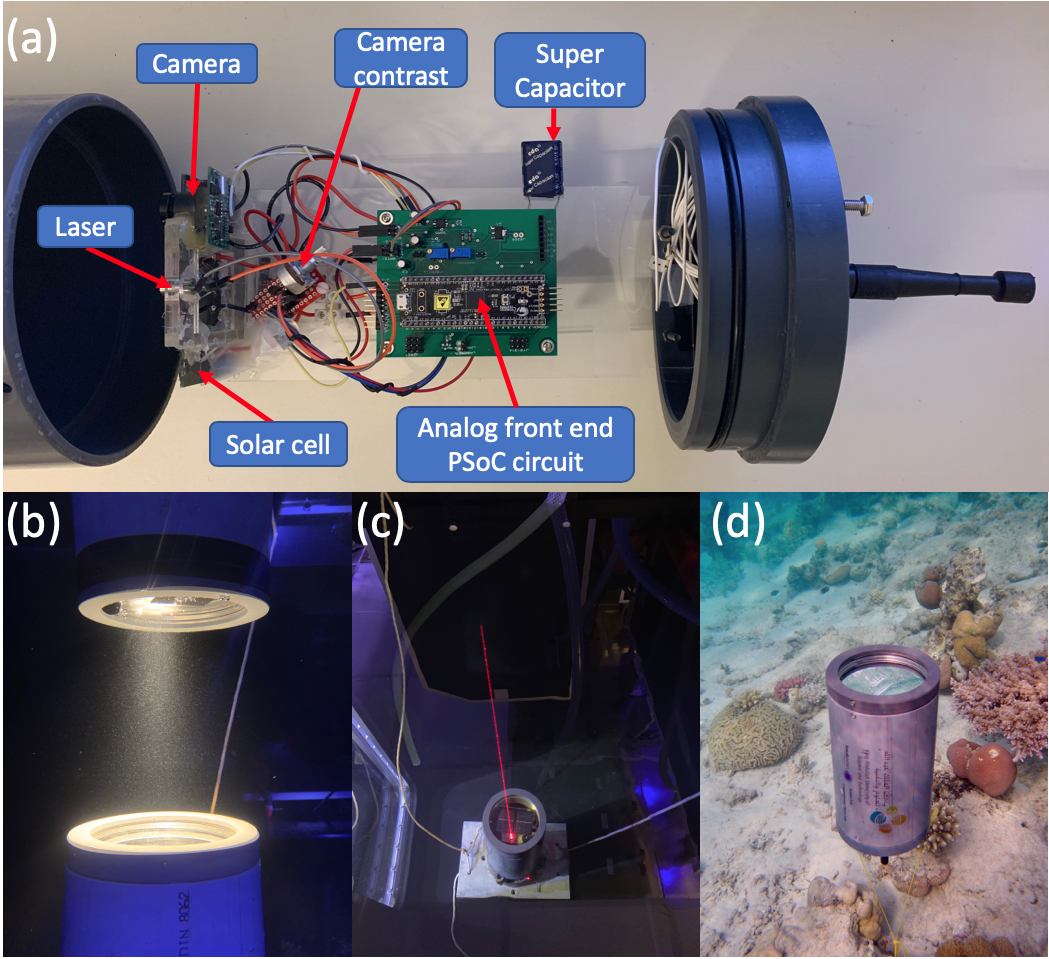}
\caption{(a) Self-powered underwater camera.  Photograph of the module (b) being charged by an LED source, (c) transmitting information (video streaming) with a red laser, and (d) deployed in a coral reef in the Red Sea.}
\label{Camera}
\end{figure}

The second demonstration involves an IoUT device equipped with an analog camera for underwater live video streaming, as shown in Fig.~\ref{Camera} (a). The IoUT is formed by an analog front-end PSoC circuit, powered by a 5 F super-capacitor, which can be charged via a solar panel (similar to the $55\times70$ mm used to perform the first demonstration). A low-power red laser is also connected to the circuit for video transmission. The device is fixed at the bottom of a vertical tank with sea water.
Using an LED source, as seen in Fig.~\ref{Camera} (b), that is fixed 30 cm away from the solar panel, the full super-capcitor charge takes approximately 1h 30min. Once fully charged, the device is used to establish a 1 minute-lomg real-time streaming of a video captured by the analog camera, as seen in Fig.~\ref{Camera} (c).\newline
The real-life deployment of the self-powered device in the Red Sea (location known as Abu Gisha island) is shown in Fig~\ref{Camera} (d). With the strong water movement that significantly shakes the device. The use of the large-area solar panel to harvest energy and decode information eased the PAT requirements compared to state-of-the art UWOC systems based on limited active area detectors. However, water motion significantly affected the pointing of the device's laser towards the detector.
\section{Open Problems}
There are different challenges associated with the various SLIPT techniques, which are either related to the hardware such as the bandwidth of the solar cell and the battery life time or the propagation effects over wireless underwater media. Here we discuss the open problems of the SLIPT technology and propose future research directions with the objective of coping with deployment challenges.
\subsection{Hardware Challenges}
One of the main challenges for SLIPT is to increase the transmission rate of optical communication links. The problem lies with the bandwidth of the receiving terminal, which is limited by the decoding bandwidth of the solar cell, mainly for time switching SLIPT configurations. The bandwidth of commercially available solar cells is usually restricted to a few tens of KHz. Using advanced modulation formats such as M-quadrature amplitude modulation-orthogonal frequency division multiplexing (M-QAM OFDM) can potentially scale the transmission capacity by several orders of magnitude. However, implementing such techniques requires the use of a digital-to-analog converter (DAC) at the transmitter and an analog-to-digital-converter (ADC) at the receiver, which comes at a major cost, which is the battery energy consumption. Limitations on data transmission rates can be alleviated in a power splitting SLIPT approach where a high-bandwidth PD can be used to decode the information signals instead of the solar detector at the expense of the pointing errors since the detection areas of commercially available high-bandwidth PDs are limited to only few tens of mm$^2$, due to the limit imposed by the resistor-capacitor (RC) time constant \cite{RCTime}, which results in a strict angle of view that requires maintaining a perfect system alignment. Using high-speed PDs can also generate additional complexity for the power splitter, which should be adapted to deliver sufficient power to decode the signals.\newline
Additional challenges are related to the life time of the IoUT device 's battery, which is subject to the enabled features and the information transmission rate. To provide an idea on the impact of enabled features and data throughput on the energy consumption levels of IoUT devices, we collected in Table \ref{CurrentConsumption} the current consumption portion of a fully awake device for various energy sources  with different characteristics  as well as data throughput levels (data decoded by the solar cell in the photoconductive mode). 
\begin{table}[!t]
\renewcommand{\arraystretch}{1.3}

\caption{Current consumption for different enabled features}
\label{CurrentConsumption}
\centering
\begin{tabular}{|l|l|l|l|}
\hline
\rowcolor[gray]{0.8}
Source&\thead{Current \\Consumption}&Throughput & Enabled Features\\
\hline
3.7 V&102 mA&500 Kbit/s&Wi-Fi, Bluetooth\\
\hline
3.7 V&36 mA&500 Kbit/s&IoT with clock at 10MHz\\
\hline
3.7 V&11 mA&115.2 Kbit/s&\thead{SoC with microcontroller\\ at 3MHz}\\
\hline
5 V&110 mA&-&Video streaming\\
\hline
3.7 V&7 mA&-&Sensing and saving data\\
\hline
5 V&236 mA&500 Kbit/s&\thead{Video streaming,\\ Wi-Fi and Bluetooth}\\
\hline
\end{tabular}
\end{table} 
\subsection{Propagation Effects}
When propagating through the water, the intensity of a light  beam decays exponentially along the propagation direction $z$, from the initial intensity $I_{0}$ following Beer's law expressed as $I=I_{0}\exp(-\alpha z)$, with $\alpha$ being the absorption coefficient. $\alpha$ is obtained by summing the contribution of two main phenomena, $b$ and $c$, which are the absorption and the scattering coefficients, respectively. Beams propagating through the water can be also subject to turbulence, which is due to random temperature in-homogeneity, salinity variations or air-bubbles. Temperature fluctuations and salinity variations result in rapid changes in the refractive index of the water, while air-bubbles block partially or completely the light beam. Statistical models to estimate the impact of turbulence on the underwater link  under several  conditions are existing in the literature \cite{UWOCModels}. When designing a link to simultaneously transfer power and information in an underwater channel, attenuation as well as turbulence effects should be taken into account  to ensure the delivery of a sufficient amount of power to perform the two SLIPT functions. \newline
A possible way to reduce the impact of turbulence is two use multiple wavelegths for the information transfer and energy harvesting functions. The use of multiple wavelengths can provide a diversity gain over a harsh underwater environment, if different copies of the same signal are encoded over distinct carrier wavelengths that are affected differently by turbulence-induced distortions \cite{Colors}.  For example, light attenuation in clear seawater is minimum in the blue-green region, while longer wavelengths are more effective to mitigate the effect of turbulence. At clear water,  the use of multiple wavelengths to carry independent data streams can considerable scale the transmission capacity. Taking the example of a two-wavelength system, one wavelength can be used to transfer energy while the other can carry the data streams ensuring a continuous connectivity of the device. The two wavelengths can be also used to transmit two independent data streams and charge the battery at the same time if a power splitting approach is adopted.
\subsection{Beam Divergence}
While propagating through an unguided medium, a light beam tends to diverge leading to an increase of the radius. Losses due to beam divergence can be denoted as geometrical attenuation, which scales with the propagation distance and related to the used laser or LED source at the transmitter (including the collimation system, if used) as well as the operating wavelength. Taking into account beam divergence is crucial for SLIPT systems. The power portion at the receiver should fulfill the two SLIPT functions.
\subsection{Path Obstructions}
\indent Obstructed propagation path is another limiting factor for underwater SLIPT, which can be overcome using the enhanced scattering of the ultraviolet (UV) light that can be harnessed to establish non-line of sight (NLoS) connections. Nonetheless, this requires having solar-blind solar cells to harvest energy from UV light. This technique should be also carefully studied to avoid UV exposure health-related issues.
\section{Conclusion}
Throughout this article, we provided an overview of the SLIPT technology. SLIPT is a key technology for green energy transfer that could exploit different degrees of freedom including time, power, and space. We presented two underwater experimental demonstrations of time switching SLIPT. In a first proof-of-concept test, we charged the battery of a submerged module using a blue laser and successfully transmitted commands with a transmission rate of 500 Kbit/s through a 1.5 m underwater link. We also collected data using the self-powered temperature sensor.  In the second experiment, we report transmitting commands and charging the capacitor of a device equipped with a low-power red laser and analog-camera. SLIPT is still a largely unexplored field and requires deeper research efforts to overcome several major technical issues and channel-related challenges before its wide-scale deployment.

\section*{Acknowledgment}
This work was supported by funding from King Abdullah University of Science and Technology.  The authors would like to thank the Red Sea Research Center and Coastal \& Marine Resources Core Lab  (CMOR) for helping in the testing, and deployment of the prototypes.
\ifCLASSOPTIONcaptionsoff
  \newpage
\fi

\begin{IEEEbiographynophoto}{Jose Ilton de O. Filho} received the B.S. degree in Electrical Engineering from Federal University of Piaui, Brazil, in 2017. He is currently pursuing the M.Sc. degree in Electro-Physics at King Abdullah University of Science and Technology (KAUST), Thuwal, Makkah Province, Saudi Arabia. His current areas of interest include instrumentation, sensing, optical communication, and wireless energy harvesting.
\end{IEEEbiographynophoto}

\begin{IEEEbiographynophoto}{Abderrahmen Trichili} received his dipl\^ome d'ing\'enieur and PhD degree in Information and Communication Technology from l'\'Ecole Sup\'erieur des Communications de Tunis (SUP'COM, Tunisia) in 2013 and 2017, respectively. He is currently a Postdoctoral Fellow in CEMSE at KAUST. His current areas of interest include space division multiplexing, orbital angular momentum multiplexing, free-space optical communication, and underwater wireless optical communication systems.
\end{IEEEbiographynophoto}

\begin{IEEEbiographynophoto}{Boon S. Ooi} is a Professor of Electrical Engineering at KAUST. He received the PhD degree from the University of Glasgow (UK) in 1994. He joined KAUST from Lehigh University (USA) in 2009. His recent research is concerned with the study of III\nobreakdash-nitride-based materials and devices and lasers for applications such as solid-state lighting, visible-light and underwater wireless optical communications, and energy-harvesting devices. He has served on the technical program committee of CLEO, IPC, ISLC, and IEDM. He currently serves on the editorial board of Optics Express and IEEE Photonics Journal. He is a Fellow of OSA, SPIE, and IoP (UK).
\end{IEEEbiographynophoto}

\begin{IEEEbiographynophoto}{Mohamed-Slim Alouini} was born in Tunis, Tunisia. He received the PhD degree in Electrical Engineering from the California Institute of Technology (Caltech), Pasadena, CA, USA, in 1998. He served as a faculty member in the University of Minnesota, Minneapolis, MN, USA, then in the Texas A\&M University at Qatar, Education City, Doha, Qatar before joining KAUST, Thuwal, Makkah Province, Saudi Arabia as a Professor of Electrical Engineering in 2009. His current research interests include the modeling, design, and performance analysis of wireless communication systems.
\end{IEEEbiographynophoto}

\begin{IEEEbiographynophoto}{Khaled N. Salama} received the B.S. degree from the Department Electronics and Communications, Cairo University, Cairo, Egypt, in 1997, and the M.S. and Ph.D. degrees from the Department of Electrical Engineering, Stanford University, Stanford, CA, USA, in 2000 and 2005, respectively. He was an Assistant Professor at Rensselaer Polytechnic Institute, NY, USA, between 2005 and 2009. He joined King Abdullah University of Science and Technology (KAUST) in January 2009, where he is now a professor, and was the founding Program Chair until August 2011. He is the director of the sensors initiative a consortium of 9 universities
(KAUST, MIT, UCLA, GATECH, Brown University, Georgia Tech, TU Delft, Swansea University, the University of Regensburg and the Australian Institute of Marine Science (AIMS). His work on CMOS sensors for molecular detection has been funded by the National Institutes of Health (NIH) and the Defense Advanced Research Projects Agency (DARPA), awarded the Stanford-Berkeley Innovators Challenge Award in biological sciences and was acquired by Lumina Inc. He is the author of 250 papers and 18 US patents on low-power mixed signal circuits for intelligent fully integrated sensors and neuromorphic circuits using memristor devices.
\end{IEEEbiographynophoto}

\end{document}